# Alloy theory with atomic resolution for Rashba or topological systems

Zhi Wang[1], Jun-Wei Luo[2,3] and Alex Zunger[1, *]

[1] *Renewable and Sustainable Energy Institute, University of Colorado, Boulder, Colorado 80309, USA*

[2] *State Key Laboratory for Superlattices and Microstructures, Institute of Semiconductors, Chinese Academy of Sciences, Beijing 100083, China*

[3] *Beijing Academy of Quantum Information Sciences, Beijing 100193, China*

*alex.zunger@colorado.edu

Interest in substitutional disordered alloys has recently reemerged with focus on the symmetry-sensitive properties in the alloy such as topological insulation and Rashba effect. A substitutional random alloy $(AX)_x(BX)_{1-x}$ of components AX and BX generally manifests a *distribution of local environments,* whereby each X site for example can be locally decorated by different substitutional arrangements of {A, B, X} atoms, thus creating an inherently *polymorphous network*. Electrons will then respond to the existence of different local environments and site symmetries, creating local charge transfer and atomic displacements patterns observed in experiments. While the macroscopic average structure $S_0$, as seen by probes with long coherence length, may have the original high symmetry of the constituent compounds, many observable physical properties are sensitive to local symmetry, and are hence the average $<P(S_i)>$ of the properties {$P(S_i)$; i=1…N} of the individual microscopic configurations {$S_i$; i=1,N} rather than the property $P(<S_i>)=P(S_0)$ of the macroscopically averaged high-symmetry (monomorphous) configuration $S_0$. The fundamental difference between the polymorphous representation $<P(S_i)>$ vs the monomorphous $P(S_0)$ in modeling substitutionally disordered alloys led to the often diverging results between methods that 'see' atomic details and those that see only the high symmetry of the constituents, while missing the atomic-scale resolution needed in many cases to discern local symmetry-related physics. A natural approach that captures the polymorphous aspect of random alloys is the well-known supercell approach where lattice sites are occupied by the alloyed elements with a particular form of disorder and solved via periodic electronic structure methods for sufficiently large supercells. However, such approaches tend to produce complex *E* vs ***k*** dispersion relations ('spaghetti bands'), rendering the wavevector *k* information needed in theories of topology and Rashba physics and seen in angular resolved experiments, practically inaccessible. The results of such calculations have consequently been displayed as density of states. A solution that retains the polymorphous nature of the random alloy but reinstates the *E* vs ***k*** relation in the base Brillouin zone is to unfold the supercell bands. This yields alloy "Effective Band Structure" (EBS), providing a three-dimensional picture of the distribution of spectral density in the whole Brillouin zone. It consists of *E*- and *k*-dependent spectral weight with coherent and incoherent features, all created naturally by the specific nature of the chemical bonding underlying the polymorphous distribution of many local environments. We illustrate this EBS approach for CdTe-HgTe, PbSe-SnSe and PbS-PbTe alloys, showing atomic-scale effects such as formation of a *distribution* of A-X and B-X bond lengths, local charge transfer, and the creation and destruction of valley degeneracies. In CdTe-HgTe the disorder effect is so weak that the incoherent term is negligible, and the monomorphous approaches are still feasible in this alloy. In PbSe-SnSe the stronger disorder effect introduces significant (~150 meV) *band splitting of the topological band inversion*, forming a *sequential* inversion of multiple bands which is important for the topological transition but absent in monomorphous methods. In PbS- PbTe there is a strong disorder effect, revealing the emergence of ferroelectricity from the polymorphous network in this alloy.

## I. Introduction

*Alloy theories with or without atomic resolution:* Many target properties of materials are not available in currently known individual components AX or BX but do exist in alloys of such components $(AX)_x(BX)_{1-x}$, where X denotes anion and *x* denotes alloy composition. Examples include band gap and effective mass tuning in semiconductors, ductility, brittleness and a given degree of short-range ordering (SRO) in intermetallics, and topological properties existing only after alloying[1]. Inevitably, disorder effects due to the substitutional occupations of A and B atoms in $(AX)_x(BX)_{1-x}$ is the key to understand alloy properties such as mobility, conductivity, electronic structure and localization. Of particular recent interest are *alloy properties that depend on local symmetry*, such as the emergence of Rashba effect, predicated on absence of inversion symmetry, in alloys of centrosymmetric components[2], or the appearance of topological properties in alloys *e.g.* $(PbSe)_x(SnSe)_{1-x}$[3] and $Mo_xW_{1-x}Te_2$[4] at specific, time reversal invariant wavevectors.

However, accounting for local symmetry effects in random disordered alloys is never a simple task. Common models of disorder have considered single-site models and continuum model that account for the changes of lattice vectors and volume (*e.g.*, via Vegard's law) but retain the macroscopic symmetry rather than the local symmetry. The Virtual Crystal Approximation[5] (VCA) relies on the assumption that alloy short-range disorder has negligible effects and can be averaged out, thus largely restoring in the alloy the symmetry (hence band structure shapes and degeneracies) of the parent compounds; the Single-site Coherent Potential Approximation[6] (S-CPA) with account k-dependent broadening of the band structure[7] approximately calculates the microscopic (local) environment however neglects effects such as local symmetry lowering due to atomic relaxation. Nevertheless, the VCA and S-CPA generally lack a full description of atomic-scale resolution of disorder that should be visible when the alloyed elements differ sufficiently on some scale of atomic sizes, bonding characteristics, or charge transfer.

The insufficiency of the monomorphous alloy description has been shown in many previous works. Examples include (i) the extended X-ray absorption fine structure (EXAFS) and atomic pair distribution function (PDF) observation of the existence in random alloy of a *distribution of A-X and B-X bond lengths*[8–13]; and (ii) the observation that atomic site charges $\{Q_i\}$ in an alloy depend on the local neighborhood environment of site *i*, which results in a non-vanishing, large electrostatic alloy (Madelung) energy[14], contradicting the common assumption underlying S-CPA, of the independence of charges on local environment, leading to $<Q_iQ_j> = <Q_i><Q_j> = 0$ *i.e.* vanishing electrostatic energy. To achieve an atomic resolution of disorder one needs theory that recognizes local symmetries, yet informs about the extent to which the long-range translational symmetry underlying the concept of wavevector is retained in alloys.

Theories of topological effects in random alloys[15–17] argue that in an infinite sample of random alloy all symmetry elements (*e.g.* inversion center) of the constituent solids being mixed are preserved *on average, so the latter configuration can be used to evaluate topological characteristics*. However, even if this proposition were correct, the properties of the alloy <P> (such as band structure and band inversion) do not reflect the property of the macroscopically averaged configuration <P>=P($S_0$) but rather the average $P_{obs}=\Sigma P(S_i)$ of the properties {P($S_i$)} of the individual microscopic configurations {$S_i$; i=1, N}.

***The atomically resolved perturbations induced by A-on-B substitution in alloys:*** As is well known**,** disorder models with atomic resolution can be built by solving the band structure problem of supercell whose $N \times N \times N$ primitive cells contain $N^3$ sites are occupied randomly by the constituent atoms *A* and *B*. The alloyed atoms can naturally have different electronic structures, atomic sizes, and tendencies for charge transfer, thereby creating a polymorphous representation where (unlike VCA or S-CPA) the common atom *X* is 'seeing' a variety of local environments, depending on the number of *A* and *B* atoms locally coordinating different *X* sites. In this representation, *A*-X, B-X and A-B charge transfer, as well as the existence of a distribution of *A-X* and *B-X* bond lengths is allowed, in addition to the trivial variation of volume with composition. The spectra can be converged with respect to the size of the supercell and by averaging over a representative number of random realizations. More effectively, one can construct from the outset special supercells, 'Special Quasirandom Structures' (SQS) that are guaranteed to reproduce pair and many body correlation functions in the best way possible for a given supercell size N[18]. The Observable property P calculated for such an SQS structure is not simply the property of a single snapshot configuration but approximates the ensemble average <P> for the random configuration. This is described in Ref[18,19]. Furthermore, in general, SQS supercell with large size gives more reliable result than the ensemble average along many small random supercells, as shown by Ref[19]. The reason is that large supercells contain intermediate range interactions (such as 4th neighbor pairs inside a supercell) which do not exist in small supercells, where such interactions are approximated by the replica of the interactions *outside* the small cell. Indeed, when a physical property needs for its description contributions that scale as n-th order pair interaction, then small cells have a limited $n_{max}$ value since further values of n > $n_{max}$ are replicas of other n and contain no new information, whereas large supercells are needed to capture longer range pairs that come from same supercell. Convergence tests to P as a function of SQS size were tested as shown in the Methods section. Details can be found in Ref[18,19].

***The limitation of supercells and band unfolding:*** As the size of supercells increase, the *E* vs *k* dispersion relation also becomes more complex because of band folding. This leads to the difficulty of interpreting alloy effects that depend on wavevector *k* information, *e.g.* topological materials, Rashba physics and Angle-Resolved Photoemission Spectroscopy (ARPES) analyses. Such effects would be concealed inside the 'spaghetti-like' supercell bands. This is perhaps the primary reason density of states, rather than *E* vs *k* dispersion, is usually shown in such supercell calculations. This difficulty can be solved by the "Effective Band Structure" (EBS) method[20], which unfolds the supercell band structures into the primitive Brillouin zone (BZ), same as the BZ used in the theoretical study of pure compounds and the experimental ARPES study of alloy. Similar to ARPES, the EBS method also provides a three-dimensional picture of the spectral function with E- and k-dependent features including coherent (dispersive term, or 'sharpness') and incoherent (band non-dispersive broadening, or 'fuzziness') spectral weights[21], all naturally produced by the polymorphous nature of the many local environment in alloys. Depending on the electronic structure method used to solve the supercell Hamiltonian (mean-field like approaches, or explicitly correlated approaches), additional coherent or incoherent effects originating from many-body effects can come in. Here we emphasize that even a single determinant electronic structure method such as Density Functional theory (DFT) will already produce three-dimensional spectral functions with coherent and incoherent features just because of allowing a polymorphous representation of the structure.

We applied our supercell model with DFT to several substitutionally random alloys CdTe-HgTe (topological alloy), PbSe-SnSe (topological alloy) and PbS-PbTe (bulk Rashba alloy). By using the EBS, we restored the *E* vs ***k*** band dispersion for alloy into the primitive BZ. We found that:

(1) In **CdTe-HgTe** is a weakly perturbed alloy made of nearly size-matched components of similar electronic and bonding properties, with band gap occurring at nondegenerate state at the Γ point. The polymorphous theory gives band structure and topological band inversion point that are rather similar to those found previously in the monomorphous theory.

(2) **PbSe-SnSe** is an alloy with moderate chemical disparity in the alloyed elements Pb vs Sn but with degenerate band edge states at L point. Here, the monomorphous theory fails to describe the disorder-induced band edge splitting, whereas the polymorphous representation shows that the states split and invert *sequentially*. This is because the monomorphous approaches do not consider the symmetry breaking induced by charge exchange and bond relaxation, masking such events by a sweeping band broadening parameter. Such approaches are inadequacy for prediction of topological properties in this system.

(3) **PbS-PbTe** represents a strongly scattering alloy (8% lattice mismatch). We find in the high-resolution picture a coherent, Rashba-like band splitting (revealing the ferroelectricity) emerging from the incoherent band broadening (revealing the alloy disorder). The mixture of coherent and incoherent features in this alloy is absent in the monomorphous approaches.

**II. Modeling the physical changes in the constituent compounds upon forming an alloy.**

To analyze the specific physical effects contributing to alloy formation we will decompose the alloy formation into three physical steps illustrated in Fig. 1:

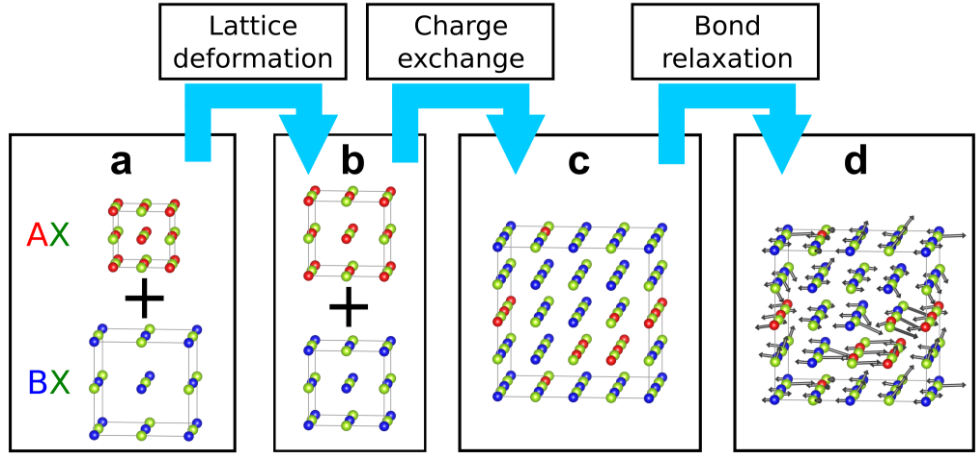

FIG. 1. The physical steps from pure compounds to alloy. (a) Pure compounds AX and BX at their equilibrium lattice vectors $\{\vec{a}_0(AX)\}$ and $\{\vec{a}_0(BX)\}$; (b) deform AX and BX lattice vectors, and let them both have the alloy lattice vectors $\{\vec{a}_0(x)\}$; (c) form a supercell of AX-BX with lattice vectors $\{\vec{a}_0(x)\}$ while keeping all bong lengths equal to each other; (d) relax all bonds in the supercell with lattice vectors $\{\vec{a}_0(x)\}$. Step a→b is the lattice deformation (LD); step b→c is charge exchange (CE), whereas step c→d is the bond relaxation (BR) step.

(1) ***Lattice deformation (LD) step*** (from Fig. 1(a) to Fig. 1(b)): Here we prepare the two constituent compounds so they could form a common alloy lattice in the next step. To do so we distort the lattice vectors for both components, so both have the same lattice vectors $\{\vec{a}_0(x)\}$ appropriate to this alloy of composition *x*, where the subscript 0 means the equilibrium lattice vector. $\{\vec{a}_0(x)\}$ are determined by doing a full energy minimization relaxation (atomic positions, cell shape and cell volume) of a large SQS supercell of that alloy at composition *x*. For the alloys of CdHgTe and PbSnSe, the Vegard lattice constant $a_{Vegard}(x)$ is very close to $a_0(x)$, while for PbSTe alloy $a_0(x)$ becomes concave above $a_{Vegard}(x)$. For example, pure PbS and PbTe are both face-center cubic (FCC) structure, with $a_0(PbS)$ = 6.03 Å, $a_0(PbTe)$ = 6.55 Å and α = β = γ = 90 degree; while in PbSTe alloy the lattice constant has changed ($a_0(PbS) < a_0(x) < a_0(PbTe)$), and the crystalline structure has transformed into a distorted rhombohedral structure. We then in this step expand the smaller component (here, PbS) and compress the larger component (here, PbTe), and distort the two compounds into the same, distorted rhombohedral structure as in alloy. Note that the alloy lattice constants as well as the cell distortions are calculated from DFT and validated with experiments. The change in extensive property P(*x*) (total energy *etc.*) in this step can be modelled by the formal reaction

$$AX|_{\vec{a}_0(AX)} + BX|_{\vec{a}_0(BX)} \rightarrow AX|_{\vec{a}_0(x)} + BX|_{\vec{a}_0(x)} \tag{1}$$

This step reveals the contribution of the deformation of lattice on the alloy formation.

(2) ***Charge exchange (CE) step*** (from Fig. 1(b) to Fig. 1(c)): Here we mix the structures prepared in the previous step to form the random alloy supercell at lattice vectors $\{\vec{a}_0(x)\}$. At this step the A-X and B-X bonds are allowed to coexist in the alloy so charge exchange can occur among different atomic sites, but all bonds are still constrained to equal to each other. Each atomic site of a given chemical identity (such as X of AX) will have in principle, a different charge distribution around it, generally reflecting the number of A and B atoms around it. CE step is a polymorphous effect thus not captured by monomorphous approaches. The change in extensive property P(x) in this step can be modelled by the formal reaction

$$AX|_{\vec{a}_0(x)} + BX|_{\vec{a}_0(x)} \rightarrow A_{1-x}B_xX|_{\vec{a}_0(x)} \qquad (2)$$

representing charge exchange at constant volume and ideal bond geometry.

(3) **Bond relaxation (BR) step** (from Fig. 1(c) to Fig. 1(d)): Here we take the previous step where a supercell with its attendant charge transfer was already formed and now allow the relaxation for all internal atomic positions at the fixed alloy lattice vectors $\{\vec{a}_0(x)\}$. Note that for each composition the bond lengths are not single-valued but have distributions due to the polymorphous local environment effect, i.e., bond lengths $R_{A-X}^{(n)}(x)$ and $R_{B-X}^{(n)}(x)$ are neighborhood-configuration-dependent (($n$)-dependent). The BR step is a polymorphous effect thus not captured by monomorphous approaches. The change in extensive property P(x) in this step can be modelled by the formal reaction

$$A_{1-x}B_xX|_{\vec{a}_0(x)} \rightarrow relaxed\ A_{1-x}B_xX|_{\vec{a}_0(x)} \qquad (3)$$

The total change in extensive property P(x) of alloy relative to the linearly weighted average of the constituents can be written as

$$\Delta P_{tot}(x) = P(x) - [xP(AX) + (1-x)P(BX)] = \Delta P_{LD}(x) + \Delta P_{CE}(x) + \Delta P_{BR}(x) \qquad (4)$$

which will assist us in analyzing physical alloy effect.

### III. Computational details

This work used the computational resources of the Extreme Science and Engineering Discovery Environment (XSEDE)[22]. We have performed DFT calculations as applied within the Vienna *ab initio* simulation package (VASP)[23] using the projector-augmented wave (PAW)[24] pseudopotentials. Cd 4$d$, 5$s$, Hg 5$d$, 6$s$, Pb 5$d$, 5$s$, 5$p$, Sn 4$d$, 5$s$, 5$p$, S 3$s$, 3$p$, Se 4$s$, 4$p$ and Te 5$s$, 5$p$ have been treated as valence electrons. For all primitive cells of pure compounds, we used an 8×8×8 Γ-center $k$ mesh in the electronic self-consistent iterations and in the atomic relaxations. Table I shows the space group, energy cutoff and exchange correlation terms, and the comparison of relaxed lattice constant and band gap for all pure compounds between DFT and experimental results. All alloy supercells have been constructed using the SQS method as implemented in the Alloy Theoretic Automated Toolkit (ATAT)[25,26]. Alloy supercell sizes are 32 formula unit (f.u.) (CdTe-HgTe), 128 f.u. (PbSe-SnSe) and 32 f.u. (PbS-PbTe), while the $k$ meshes are 4×4×4 (CdTe-HgTe), 3×3×2 (PbSe-SnSe) and 4×4×4 (PbS-PbTe). We calculated all alloy supercells using the same parameters as in their constituent compounds (*e.g.*, for all CdTe-HgTe alloy supercells we used the same parameters as in CdTe and HgTe). Note that the space groups of alloys are always different from constituent compounds, because all atomic positions as well as the lattice vectors in alloys have been determined by fully relaxing (atomic positions, cell shape and cell volume) the alloy supercells. CdTe-HgTe and PbSe-SnSe alloy supercells are still in cubic phase after relaxation, however they are no longer F-43m or Fm-3m because of the polymorphous network (different atomic sites have different element occupations and different atomic displacements). This makes it completely different with the monomorphous approaches. Meanwhile PbS-PbTe alloy supercells become distorted rhombohedral after relaxation. EBS calculations have been done by a modified version of BandUP code[27].

TABLE I. The calculation details for pure compounds, and the comparison between DFT and experimental results.

| Compounds (space group) | Exchange correlation term | Cutoff energy | Lattice constant (DFT) | Lattice constant (Exp.) | Band gap (DFT) | Band gap (Exp.) |
|---|---|---|---|---|---|---|
| CdTe (F-43m) | LDA+U ($U_d$=10 eV) | 400 eV | 6.410 Å | 6.48 Å | 0.86 eV | 1.65 eV |
| HgTe (F-43m) | LDA+U ($U_d$=10 eV) | 400 eV | 6.436 Å | 6.46 Å | -0.26 eV | -0.3 eV |
| PbSe (Fm-3m) | GGA+U ($U_{Pb\_s}$=2 eV) | 360 eV | 6.22 Å | 6.12 Å | 0.23 eV | 0.17 eV |
| SnSe (Fm-3m) | GGA+U ($U_{Pb\_s}$=2 eV) | 360 eV | 5.99 Å | 6.00 Å | 0.72 eV | 0.62~0.72 eV |
| PbS (Fm-3m) | GGA+U ($U_{Pb\_s}$=2 eV) | 360 eV | 6.03 Å | 5.93 Å | 0.3 V | 0.29 eV |
| PbTe (Fm-3m) | GGA+U ($U_{Pb\_s}$=2 eV) | 360 eV | 6.55 Å | 6.44 Å | 0.2 eV | 0.19 eV |

## IV. Unfolding the supercell energy bands and recovering *E* vs *k* alloy EBS

Here we briefly summarize the basic equations of EBS. In the supercell Brillouin zone $|Km\rangle$ is the *m*-th electronic eigen state at **K**, whereas in the primitive Brillouin zone, $|k_i n\rangle$ is the n-th electronic eigen state at $\mathbf{k}_i$. Each eigenfunction $|Km\rangle$ in the supercell can be quantified by expanding it in a complete set of Bloch eigenfunctions $|k_i n\rangle$ of primitive cell, where **K** = **k**$_i$ - **G**$_i$, and **G**$_i$ being reciprocal lattice vectors in the supercell BZ. The band folding mechanism between supercell and primitive cell can then be expressed as

$$|Km\rangle = \sum_{i=1}^{N_K} \sum_n F(\mathbf{k}_i, n; \mathbf{K}, m) |\mathbf{k}_i n\rangle \quad (5)$$

where $|Km\rangle$ is the *m*-th electronic state at **K** in supercell Brillouin zone, $|k_i n\rangle$ is the *n*-th electronic state at $k_i$ in primitive Brillouin zone. One can then unfold the supercell band structure by calculating the spectral weight $P_{Km}(\mathbf{k}_i)$ from

$$P_{Km}(\mathbf{k}_i) = \sum_n |\langle Km|\mathbf{k}_i n\rangle|^2 \quad (6)$$

which is the Bloch 'preservation' of Bloch wavevector $k_i$ in $|Km\rangle$ when $E_n = E_m$. Finally, the EBS can be obtained using the spectral function $A(\mathbf{k}_i, E)$,

$$A(\mathbf{k}_i, E) = \sum_m P_{Km}(\mathbf{k}_i) \delta(E_m - E) \quad (7)$$

As an example, Fig. 2 shows the comparison among pure PbTe band structure, supercell PbSTe band structure and supercell PbSTe EBS, all plotted in the PbTe primitive Brillouin zone. The spectral function can be sharply dispersive (coherent) *e.g.* conduction band minimum (CBM) along Γ-Z direction, or become completely non-dispersive (incoherent) *e.g.* valence band maximum (VBM) at Γ, or be a mixture of both *e.g.* VBM along Γ-Z direction.

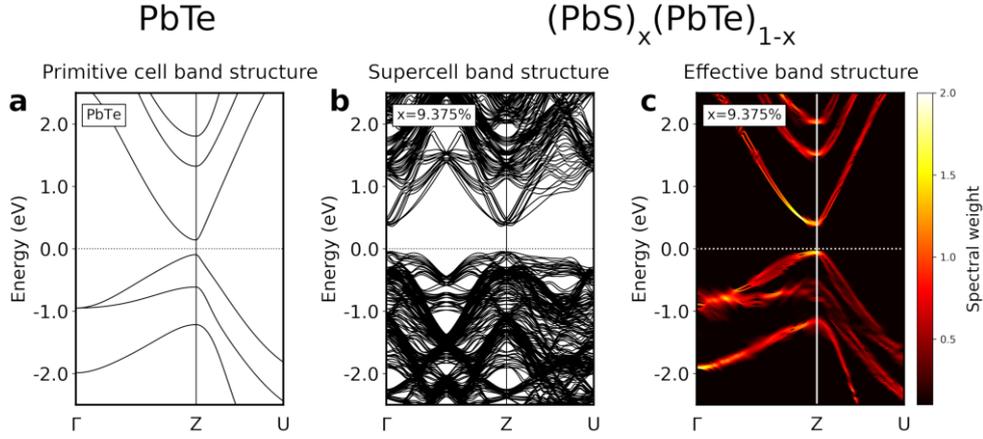

FIG. 2. (a) Conventional band structure of the distorted R3m PbTe (2-atom primitive cell, plotted in primitive Brillouin zone). (b) Supercell band structure (64-atom supercell $PbS_xTe_{1-x}$ at $x=9.375\%$, plotted in the same primitive Brillouin zone) and (c) EBS (unfolded from 64-atom supercell for $PbS_xTe_{1-x}$ alloy at $x=9.375\%$ into the same primitive Brillouin zone). The primitive lattice vectors have been distorted according to the relaxed supercell structure in order to make a direct comparison. (a) (b) and (c) are all plotted along the same Γ-Z-U direction in the primitive Brillouin zone [Z=$(\pi/a_1, \pi/a_2, \pi/a_3)$, U=$(\pi/2a_1, 2\pi/a_2, \pi/2a_3)$].

## V. Results and discussion

### A. Decomposition of alloy effects into physical terms

We study alloys having different scales of disorder: HgTe-CdTe, PbSe-SnSe and PbS-PbTe. We will see that the scale of disorder is system-dependent, from weak (HgTe-CdTe) to intermediate (PbSe-SnSe), and to strong (PbS-PbTe). It is an interesting question how different scales of disorder in different materials can affect $E$ vs $\mathbf{k}$ structure.

Fig. 3 shows the polymorphous local environment effects in the CE and BR terms in the alloy forming reactions. The effects of CE step have been shown in Fig 3(a)(c)(e) by plotting the contours of the charge density nearby one common atom in the three alloy systems. We see that the charge density around the common atom (Te in CdHgTe; Se in PbSnSe and Pb in PbSTe) has different shapes when considering the bonds formed with the dissimilar alloyed atoms, i.e., the densities around different $A$ atoms are different depending on the neighbors of $A$. The effects in BR step have been shown in Fig 3(b)(d)(f) by the bond length distribution profiles. Note that the range of y-axis becomes larger from Fig. 3b to 3f. The asymmetricity of charge density along different bonds, as well as the spread of bond length variations of different bonds, becomes more significant as one progresses from weak to strong alloying. In CdTe-HgTe, the electron density distributions along Cd-Te and Hg-Te bonds show only small differences (Fig. 2(a)), and the bond lengths of Cd-Te (as well as Hg-Te) are virtually equal with negligible distribution of values ($\sigma<0.002$ Å as shown in Fig. 2(b)). In the strongly perturbed alloy PbS-PbTe, (i) the Pb-S and Pb-Te bonds have distinct charge densities,

(ii) the bond lengths show a significant statistical spread for the same chemical bond (Fig. 2(e)(f)), *e.g.*, the Pb-S bonds vary in a range of ~1 Å, and (iii) chemically different bonds Pb-S and Pb-Te have different lengths away from the macroscopic lattice constant. Clearly, the high symmetric alloy model assumed in monomorphous theories ignore the local atomic environment effects for both electron density and geometric bond structures.

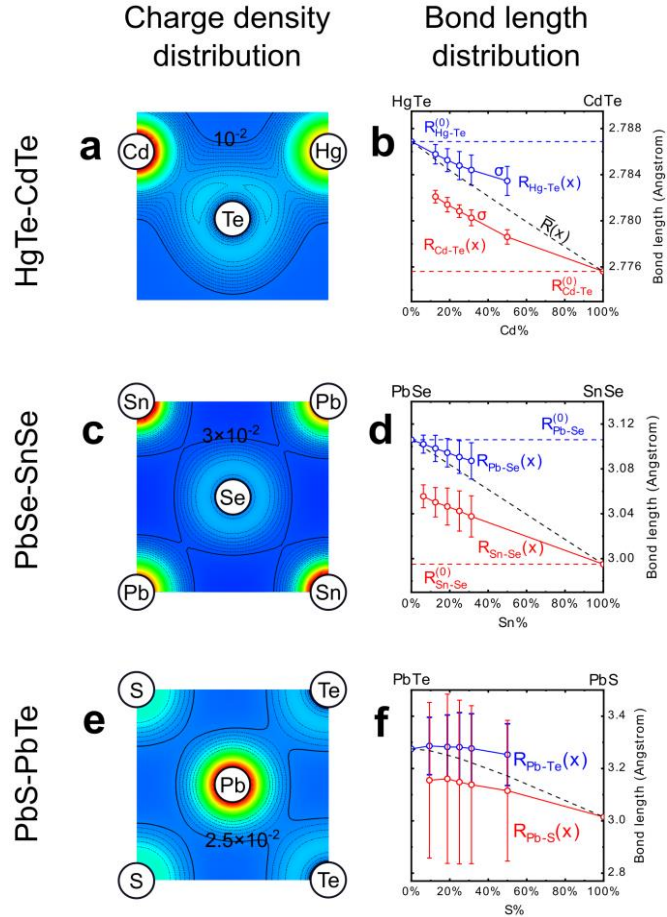

FIG. 3. The charge density and bond length profiles in three alloy systems. (a) (c) (e) show the logarithmically spaced contours for the charge density nearby one common atom for each of the alloys in CE step. (b) (d) (f) show the bond length distributions for the different types of bond (red and blue solid lines) with means (circles) and standard deviations (bars). The uniform bond lengths $R_0(x)$ in the unrelaxed lattice (*i.e.* before BR step) are shown as the black dash lines in (b)(d)(f). $R^{(0)}$ shown in red and blue dash lines are the bond lengths in pure compounds.

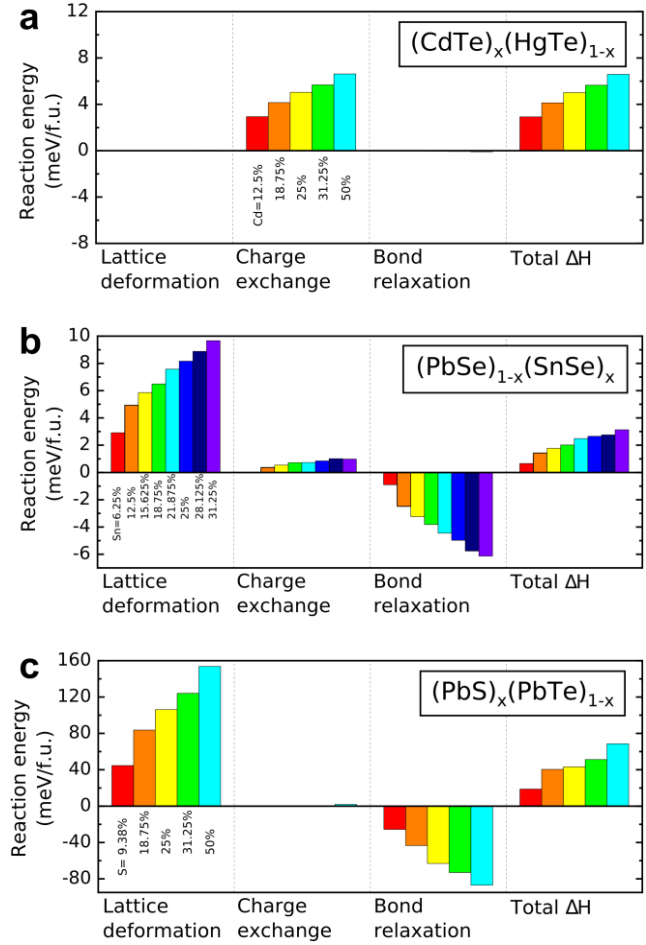

FIG. 4. Alloy mixing energies in each physical step and the total formation energies, for (a) CdTe-HgTe, (b) PbSe-SnSe and (c) PbS-PbTe.

The disorder effects can also be seen from the decomposition of the alloy mixing energy (total energy of alloy with respect to equivalent amounts of its constituents) from Eq. (4), shown in Fig. 4. Recall that the LD step introduces only monomorphous effects while the CE and BR steps result in polymorphous local environments. The scale of the total mixing energies in parts (a) (b) and (c) already disclose the scale of disorder, CdTe-HgTe having 7 meV ($x$=50%), PbSe-SnSe having 3 meV ($x$=31.25%), whereas PbS-PbTe having 70 meV ($x$=50%) which agrees with previous works[28,29]. In the weak alloy CdTe-HgTe where the lattice mismatch is tiny (~0.3%), the lattice deformation and bond relaxation energies are negligible, and the charge exchange contributes most to the mixing enthalpy (ΔH). As the lattice mismatch increases (PbSe-SnSe and PbS-PbTe), the bond relaxation BR step does not release as much energy as the lattice deformation costs in the first place, and the charge exchange energy is small, so the total mixing energy is positive and non-negligible. It is hence inadequate to include only the trivial lattice deformation effect as in simple monomorphous models, since the polymorphous terms, CE and BR can be important, neglecting which would lead incorrectly to high mixing enthalpies.

**B. Effective Band Structure of the three alloys:**

Here we show EBS pictures for each of the 3 alloys. We found that the spectral functions show a clear trend with respect to the scale of disorder: from weak to strong alloy, the spectral weights lose the coherent dispersion more quickly when leaving the reciprocal high-symmetric *k* points. Moreover, each alloy shows some unique features:

(1) In pure CdTe and HgTe, each band at Γ point is a twofold degenerate. CdTe-HgTe EBS shows very sharp band structure near the time reversal invariant momentum (TRIM) Γ point and no band splitting at Γ point (Fig. 5), which can be attributed to the very weak alloy disorder effect as shown in Fig 3(a)(b) and Fig 4(a). All bands near the Fermi level, including the light electron, light hole and heavy hole states (corresponding to $\Gamma_6$ and $\Gamma_8$ states in pure HgTe and CdTe), are sharply dispersive and almost 100% coherent. Note that there is a tiny splitting (<25 meV) on the heavy hole state along Γ-L direction, which is attributed to the small atomic displacement (see Fig. 3(b)) and agrees with previous work[30]. Therefore, in this alloy, the monomorphous theories might be adequate, *e.g.*, for predicting the topological band inversion between $\Gamma_6$ and $\Gamma_8$ states at Γ point as Cd composition increases.

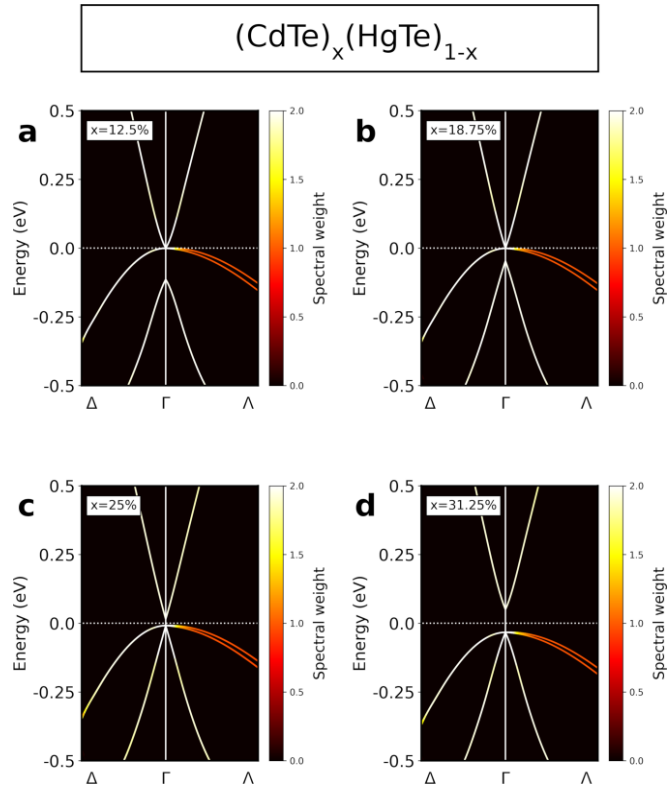

FIG. 5. EBS of CdTe-HgTe supercell (32 f.u.) at (a) Cd=12.5%, (b) Cd=18.75%, (c) Cd=25% and (d) Cd=31.25%, unfolded to the zinc-blende primitive Brillouin zone. All EBS are plotted along the same X-(Δ)-Γ-(Λ)-L direction in the primitive Brillouin zone.

(2) As the scale of disorder increasing, in PbSe-SnSe EBS, the degeneracy of band valleys at the TRIM L point is no longer preserved: there is a significant band splitting (~150 meV) on the band valleys at L point (marked by red circles in Fig. 6). Both valence band and conduction bands are split but sharply dispersive near L; moving along the Γ-L-W lines, one can see that the VBM quickly loses its coherent feature, first in the middle of Γ-L line then near the W point. More importantly, the band valley splitting at TRIM L further indicates that the 'sharp', concurrent NI-TCI transition at specific composition, which has been predicted earlier[3,7] to be due to the band inversion between highly degenerated bands, is not what a theory with atomic resolution finds: the alloy system actually experiences a regime having sequential inversions of multiple bands nearby Fermi level at L point. The sequential band inversion regime is visible only within the polymorphous model *e.g.* supercell but not in the monomorphous models[3,7,17,31–34]. Furthermore, we have found recently that the sequential band inversion not only invalidates the topological invariant of TCI-ness in such a sequential inversion regime, but also introduces a new Weyl semimetal phase between the NI-TCI phase transition, which is also absent in monomorphous alloy theories. The appearance of such Weyl semimetal phase is verified by the calculations of topological invariant and Weyl points in the supercell Brillouin zone, and can be attributed to the removal of valley-degeneracy at L point (shown in Figure 6) and spin-degeneracy nearby L point (breaking of inversion symmetry). This discussion is outside the scope of the current paper and will be discussed in a future publication dedicated to topological invariants in a random alloy.

Experimental probing of the insulator to metal transition in PbSe-SnSe alloys were carried out mostly optically. Alloy compositions where the gap is positive (insulating) were observed[35] for x < 10%, and alloy compositions where gap is smaller than 50 meV were reported[35] for 13% < x < 24%. Our calculation finds a clear insulator to metal transition, occurring in a composition regime of 12% < x < 30%. Experimentally the precise transition could not be found with IR detectors used at the time of the experiment, because gap occurs in far IR when it is smaller than 50 meV. In addition, non-randomness (i.e., clustering) and high carrier concentration can cloud the precise value of composition where the transition occurs. Perhaps a future verification of the bulk gap closing composition could be performed with low temperature, THz range optical experiment. We also hope that a verification of band edge splitting can be done in ARPES.

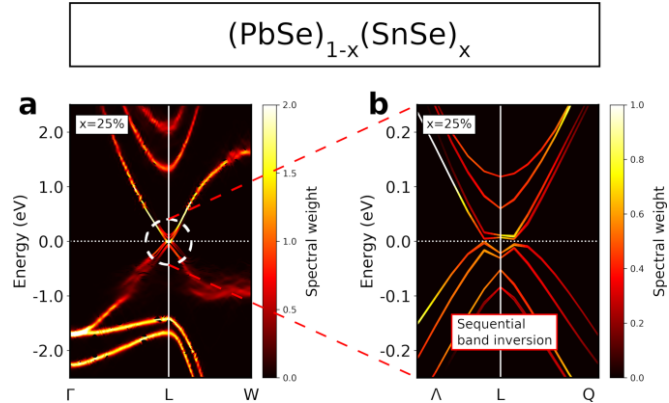

FIG. 6. EBS of PbSe-SnSe supercell (128 f.u.) at Sn=25% unfolded to the FCC PbSe primitive Brillouin zone, plotted (a) along Γ-(Λ)-L-(Q)-W direction and then (b) zoomed-in around L point. The white circles mark the sequential band inversion in this alloy attributed to band edge splitting at L point.

(3) In the strong alloy PbS-PbTe, the introduction of S atom leads to a ferroelectric (FE) sublattice displacement at low T, making the alloy a famous candidate of bulk Rashba and FE materials[10]. In VCA and S-CPA the ferroelectricity was mimic monomorphously by using a uniform displacement between cation and anion sublattice, while in our supercell, ferroelectricity is polymorphous. We found that (Fig. 7) in this alloy most bands suffer splitting and broadening, but the CBM and VBM at Z point is relatively sharp and dispersive. VCA results were previously shown in Ref.[7] Supplementary Materials Fig. 4, while CPA shown were shown in Ref.[7] Fig. 1 and Supplementary Materials Fig. 3. Comparison with our EBS results (Fig. 7) shows that the VCA is very different (no removal of degeneracies) whereas the CPA has similar band shapes as the EBS, the latter presents far more details than captured by the CPA: we see that each conduction bands clearly split into two bands along Γ-Z and Z-U directions, while VBM, although mixed with incoherent broadening, also shows such two-band splitting along Z-U direction. Note that this type of band splitting is coherent because each split branch shows an individual dispersion. We suggest that this band splitting of band edge states can be Rashba-like and reveal the ferroelectricity of this alloy system.

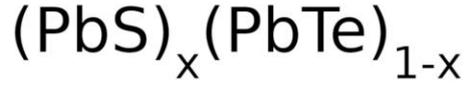

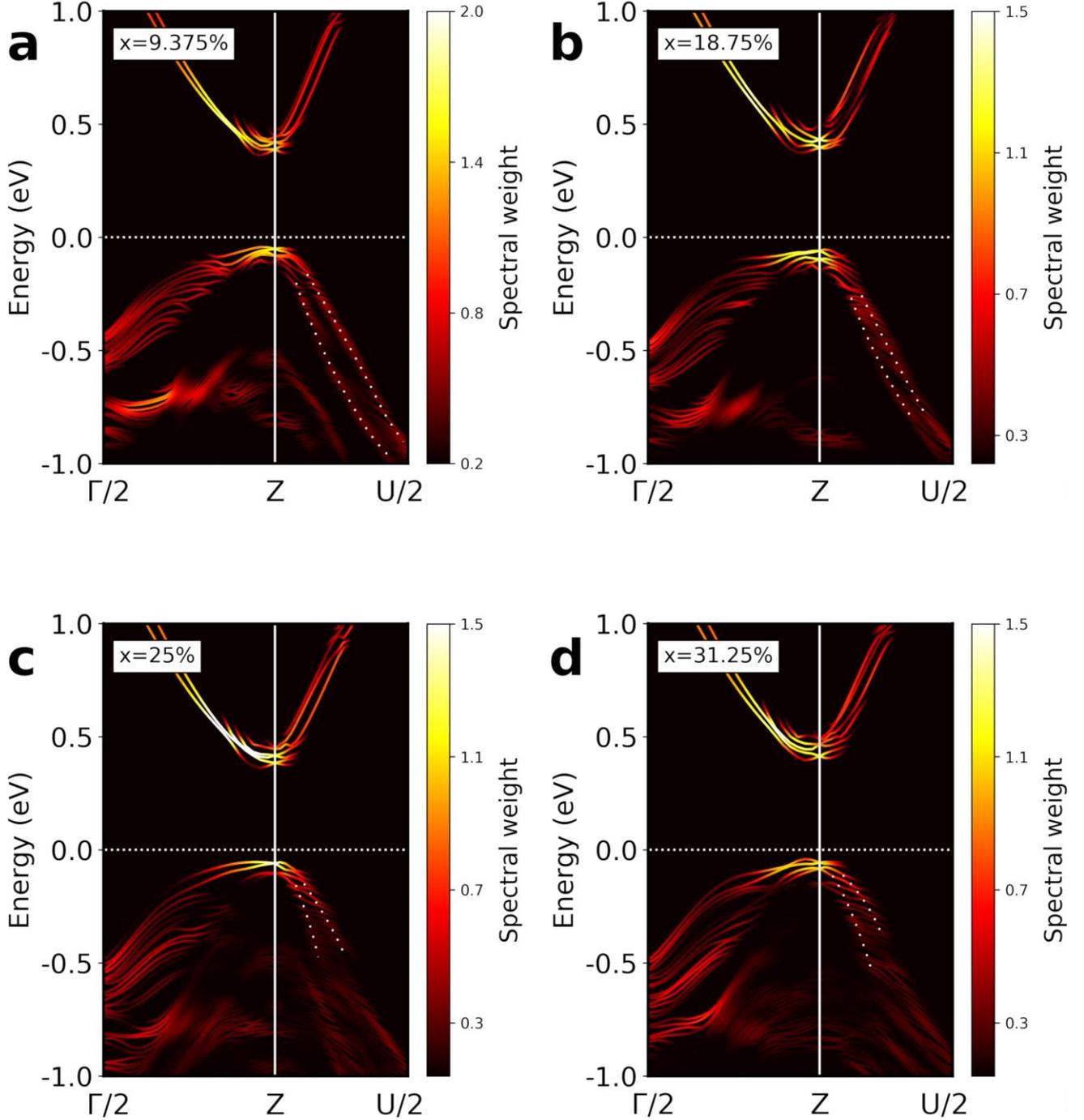

FIG. 7. EBS of PbS-PbTe supercell (32 f.u.) at (a) S=9.375%, (b) S=18.75%, (c) S=25% and (d) S=31.25%, unfolded to the distorted R3m PbTe primitive Brillouin zone. All EBS are plotted along the same Γ-Z-U direction in the primitive Brillouin zone [Z=($\pi/a_1$, $\pi/a_2$, $\pi/a_3$), U=($\pi/2a_1$, $2\pi/a_2$, $\pi/2a_3$)]. The white dot lines are only for eye-guiding to show the coherent splitting on VBM along Z-U.

## VI. Conclusions

With the aid of polymorphous supercell approach and band unfolding, we restore the all-important $E$ vs $k$ dispersion relation to alloy theory in CdTe-HgTe, PbSe-SnSe and PbS-PbTe alloys, revealing various sources of alloy formation, such as lattice deformation, charge exchange and bond relaxation. This allows one to define a scale of disorder, by the deviations that these effects create relative to the monomorphous level. We find that the spectral weights unfolded to primitive Brillouin zone shows: (1) both coherent, dispersive splitting of band degeneracies and incoherent band broadening that depends on the wavevectors and on the scale of alloy disorder; (2) coherent-incoherent transition on different bands along different k space directions; and (3) Rashba-like band splitting consisting of both coherent and incoherent features. We expect that such effects—notably the splitting of band degeneracies—could be observed by ARPES.

**Acknowledgments**

The work at the University of Colorado at Boulder was supported by the National Science foundation NSF Grant NSF-DMR-CMMT No. DMR-1724791. J.W.L. was supported by the National Natural Science Foundation of China under Grants 61888102 and 61811530022. We thank Qihang Liu for fruitful discussions on the subject. The ab-initio calculations were done using the Extreme Science and Engineering Discovery Environment (XSEDE), which is supported by National Science Foundation grant number ACI-1548562.

## Appendix: Spectral functions in ARPES and in EBS

When explaining the spectral function of primitive Brillouin zone in ARPES, a common method is to assume the outcoming photoelectron can be described by a single planewave $e^{ik\cdot r}$, i.e., a free-electron final state, therefore the spectral function can be written as[36,37]

$$\tilde{A}(\bm{k}, h\nu) = \sum_m^{occ} \sum_K^{BZ} |\bm{P}\cdot \bm{k}|^2 |\langle e^{ik\cdot r}|Km\rangle|^2 \delta(E_m - E_k + h\nu) \qquad (8)$$

where $|Km\rangle$ is the $m$-th electronic state with energy $E_m$ at $\bm{K}$ in Brillouin zone of measured sample, $h\nu$ is incoming photon energy, $E_k$ is the kinetic energy of $e^{ik\cdot r}$, and the term $|\bm{P}\cdot\bm{k}|^2$ is called the matrix element effect. $\tilde{A}(\bm{k}, E)$ in Eq. (8) represents how much *wave-vector character* of $\bm{k}$ is lost or preserved in $|Km\rangle$ when $E_k = E_m + h\nu$.

EBS, meanwhile, offers another way to calculate the spectral function: instead of single planewave, one can use the Bloch function in primitive cell $|kn\rangle$ as the final state, i.e., one calculates spectral function $A(\bm{k}, E)$ from $|\langle kn|Km\rangle|^2$ instead of $|\langle e^{ik\cdot r}|Km\rangle|^2$ as shown in Eq.(5)-(7), which is the basic concept of EBS that we describe in section IV. The EBS spectral function $A(\bm{k}, E)$ from Eq. (7) also represents the $\bm{k}$-character in $|Km\rangle$, meaning that it is comparable to $\tilde{A}(\bm{k}, E)$ in Eq. (8).

Under the single planewave final state assumption (Eq. 8), it has been proved that[36] the spectral function $\tilde{A}(\bm{k}, E)$ can be different at the equivalent $\bm{k}$ points in different Brillouin zones, e.g., first

Brillouin zone and extended Brillouin zone, even when omitting the matrix element effect. However, because the final state is Bloch function, $A(\mathbf{k}, E)$ from Eq. (7) has to obey the Bloch theorem, thus $A(\mathbf{k}, E)$ is always the same at equivalent $k$ points in different Brillouin zones. As an example, in Fig. A1 we show the EBS of $Pb_{0.75}Sn_{0.25}Se$ 256-atom supercell along the first and extended Brillouin zones: $\Gamma$ is in 1st BZ, $L_0 = (0.5, 0.5, 0.5)$ is on the boundary of 1st and 2nd BZs, while $\Gamma_2 = (1, 1, 1)$ (all have unit of $2\pi/a$) is in the extended BZ. The boundaries of first and extended BZs have been shown by white solid line. It can be seen that the intensities are the same for equivalent $k$ points in first and extended zones (same intensity along $\Gamma$-$L_0$ and $\Gamma_2$-$L_0$).

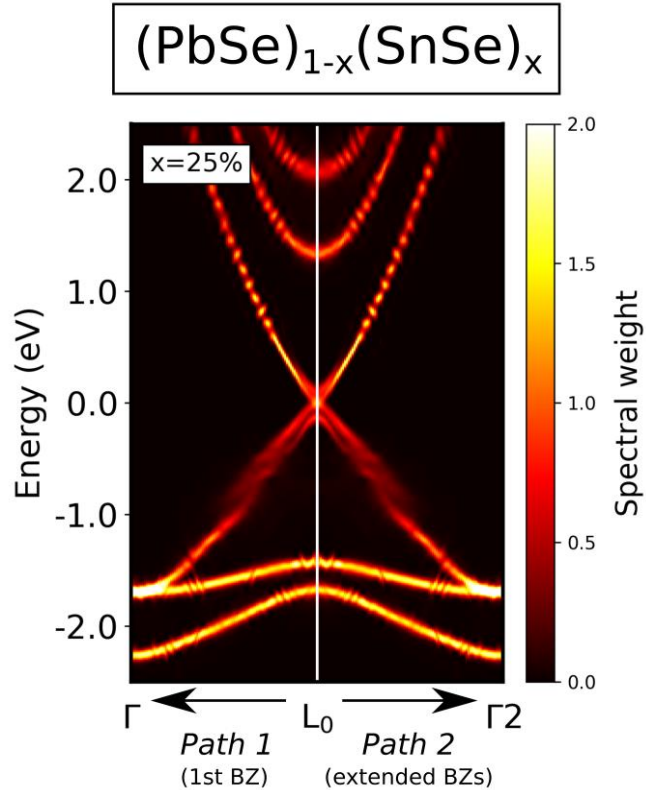

FIG. A1. EBS in a $Pb_{0.75}Sn_{0.25}Se$ 256-atom supercell. $\Gamma = (0, 0, 0)$, $L_0 = (0.5, 0.5, 0.5)$ and $\Gamma_2 = (1, 1, 1)$ (unit of length: $2\pi/a$). The white solid line at $L_0$ marks the boundary of first and extended BZs. EBS shows the same intensity along path1 (first BZ) and path2 (extended BZs).